\documentclass[referee]{raa}            

\usepackage{graphicx,times}             
\usepackage{natbib}
\usepackage{amssymb,amsmath}
\bibpunct{(}{)}{;}{a}{}{,}

\usepackage[pagebackref=true]{hyperref}
\usepackage{longtable}

\usepackage{booktabs}
\usepackage{lscape}
\usepackage{pdflscape}
\usepackage{soul}

\newcommand{\thco}{$^{13}$CO}

\newcommand{\vel}{km\,s$^{-1}$}

\newcommand{\egcite}{\citep[e.g.,][]}

\newcommand{\htcop}{H$^{13}$CO$^{+}$}

\newcommand\avir{{$\alpha_{\rm vir}$}}

\newcommand{\nth}{N$_{2}$H$^{+}$}

\begin{document}

  \title{What role of gravity, turbulence and magnetic fields in high-mass star formation clouds?}

   \volnopage{Vol.0 (20xx) No.0, 000--000}      
   \setcounter{page}{1}          

   \author{An-Xu Luo  
   \inst{1,3}
   \and Hong-Li Liu \inst{1,3}{$^{\star}$}
   \and Guang-Xing Li \inst{2}
   \and Sirong Pan \inst{1}
   \and Dong-Ting Yang \inst{1}
   }

   \institute{School of physics and astronomy, Yunnan University, Kunming, 650091, PR China; 
             {\it hongliliu2012@gmail.com}\\ 
    \and{South-Western Institute for Astronomy Research, Yunnan University, Kunming, 650091, PR China}\\
        \and{Both authors contributed equally to this work.};\\
\vs\no
   {\small Received 20xx month day; accepted 2024 Apr 11}}

\abstract{ 
To explore the potential role of gravity, turbulence and magnetic fields in high-mass star formation in molecular clouds, this study revisits the velocity dispersion--size ($\sigma$--$L$) and density--size ($\rho$--$L$) scalings and the associated turbulent energy spectrum using an extensive data sample. The sample includes various hierarchical density structures in high-mass star formation clouds, across scales of 0.01 to 100\,pc. We observe $\sigma \propto L^{0.26}$ and $\rho \propto L^{-1.54}$ scalings, converging toward a virial equilibrium state. A nearly flat virial parameter--mass ({\avir}$-M$) distribution is seen across all density scales, with {\avir} values centered around unity, suggesting a global equilibrium maintained by the interplay between gravity and turbulence across multiple scales. Our turbulent energy spectrum ($E(k)$) analysis, based on the $\sigma$--$L$ and $\rho$--$L$ scalings, yields a characteristic $E(k) \propto k^{-1.52}$. These findings indicate the potential significance of gravity, turbulence, and possibly magnetic fields all in regulating  dynamics of molecular clouds and high-mass star formation therein.
}

   \authorrunning{An-Xu Luo et al.}            
   \titlerunning{What regulates high-mass star formation clouds}  

   \maketitle
%
%
\section{Introduction}           
\label{sect:intro}
High mass stars ($M_* > 8 M_\odot$) play an important role in the evolution of galaxies and the circulation of interstellar medium. However, the process of high mass star formation remains mysterious \egcite{Mot18}. Molecular clouds (MCs) are believed to be the cradle of star formation, the study of their density substructures, and associated dynamics could help understand how high mass stars form therein \egcite{San19, Liu22a, Liu22b, Sah22,Liu23, Xu23, Hac23, Per23, Yan23, He23, Pan24}.

Molecular clouds are characterized by intricate, hierarchical density structures such as filaments, clumps, and cores. The dynamics of these hierarchical structures could be governed by competition between turbulence, gravity, and magnetic fields \egcite{Mot18, Vaz19, Bal20, Hac23}. However, the relative importance among these competing factors remains elusive in regulating dynamics of MCs and even high-mass star formation therein.
For example, there are two recent models,  'global hierarchical collapse' (GHC) model \citep{Vaz19} and the 'inertial inflow' (I2) model \citep{Pad20}, which attempt to explain the origin and evolution of massive stars and their associated structures, such as filaments, clumps, cores, and disks. Both models are based on the idea that star formation especially for high-mass star formation is a multiscale process, involving the interaction of turbulence, gravity, and feedback in a hierarchical manner. However, they differ in some aspects, such as the role of turbulence, the initial conditions, and the driving mechanisms of gas infall or mass accretion. The GHC model assumes that the initial molecular cloud is in an approximate virial equilibrium state, with turbulence being either subsonic or transonic \egcite{Vaz19}. Additionally, it posits that 
the gas infall motions at all scales are primarily driven by the self-gravity of cloud hierarchical structures. On the contrary, the I2 model contends that turbulence prevails at larger scales in molecular clouds \citep{Pad20}. It assumes that the initial molecular cloud is highly turbulent and far from equilibrium. In this scenario, turbulence is supersonic, and large-scale gas infall motions are primarily driven by the inertial inflow of turbulent gas. For the role of magnetic fields in regulating high-mass star formation and its associated dynamics, there is also a dispute between strong-field and weak-field models. The strong-field models \egcite{Mou91, Mou99} argue that the magnetic fields control evolution of the MCs and star formation within, while the weak-field models \egcite{Pad99, Mac04} favors the key role of turbulence.

From an observational point of view, the relative importance between major competitors such as gravity and turbulence, could be investigated through the empirical scaling relations of velocity dispersion and density. Those empirical scalings are often referred to as Larson's scaling relations \citep{Lar81}:
\begin{align}
\begin{split}
\left\{\begin{array}{cc}
     & \sigma \propto L^{\beta} \\
     & \rho \propto L^{-p}, \label{Larson Eq}
\end{array}
\right.
\end{split}
\end{align}
where $\sigma$ and $\rho$ are the velocity dispersion, and mass density, respectively, both measured within the size, $L$, of cloud density structures;  $\beta$ and $p$ are the scaling exponents. The first relations found by Larson yielded $\beta=0.38$ and $p=1.1$ \citep{Lar81}, later $\beta$ was refined to be 0.5 \egcite{Hey04}.
Larson’s scaling relations have been extensively studied in star formation clouds \egcite{Sol87, She12}, particularly for high-mass star formation \egcite{Cas95, Tra18, Li23}. For instance, \citet{Per23} analyzed the Larson’s scaling relations in 27 infrared dark clouds associated with high-mass star formation. They found that gravity could dominate the cloud dynamics across various scales, up to tens of parsecs. However, other observations toward high-mass star forming regions suggest that turbulence, in addition to gravity, could play a significant role in regulating the hierarchical density structures of MCs (e.g., \citealt{Hey09}, \citealt{Liu22b}, and \citealt{Pan24}). Therefore, the relative importance of these competing factors in MCs remains a topic of ongoing discussion.  

This paper aims to  revisit the Larson-like scalings, namely, the scalings of both velocity dispersion and density, providing insights onto the relative significance of gravity, turbulence, and magnetic fields in governing the dynamics of MCs and high-mass star formation therein. To this end, we collected a large sample of observational data from literature across various density scales, from giant molecular clouds to massive cores. The paper is organised as below. In Sect.\,\ref{Sec2}, we depict the theoretical derivations of general Larson-like scaling relations. In Sect.\,\ref{sec:datasets}, we describe the data set. In Sect.\,\ref{sec: virialized analysis}, we primarily focus on the interpretation of the
observed Larson-like scalings and their associated energy spectrum of turbulence. In Sect.\,\ref{sec:conclusions}, we give a summary along with conclusions.

\section{Theoretical derivations for scaling relations}\label{Sec2}
For a self-gravitating object of the mass $M$ and the size $L$, being in virial equilibrium, the virial velocity is given by $\sigma_{vir} \propto (M/L)^{1/2}$. According to the \cite{Li17}, the energy dissipation rate of the virial velocity is
\begin{align}
    \epsilon_{\mathrm{vir}} \propto M^{3/2} L^{-5/2} .
\end{align}
Under the assumption of spherical symmetry, where the mass $M \propto \rho L^3$, we find
\begin{align}
    \epsilon_{\mathrm{vir}} \propto \rho^{3/2} L^{2} .
\end{align}
The  dissipation rate of turbulent energy in the medium,  as described in \citet{Kol41}, is given by
\begin{align}
    \epsilon_{\mathrm{turb}} \propto \sigma^{3} L^{-1}.
\end{align}

If a molecular cloud maintains virial equilibrium across all hierarchical density scales,  the  transfer rate of both types of energy will be roughly equal \citep{Li17,Li18, Vaz23}:
\begin{align}
    \epsilon_{\mathrm{vir}} \simeq \epsilon_{\mathrm{turb}}.
\end{align}
From this, we obtain the relation
\begin{align}
    \rho^{3/2} L^{2} \propto \sigma^{3} L^{-1}.
\end{align}
If MCs follow the following density profile,
\begin{align}
\rho \propto L^{-p}, \label{vir_eq0}
\end{align}
we can express the turbulent velocity (\(\sigma\), corresponding to the observable of velocity dispersion) as below,
\begin{align}
    \sigma \propto L^{1-p/2}.\label{virial v equation 0}
\end{align}

Here, Eqs.\,\ref{vir_eq0} and \ref{virial v equation 0} represent the general Larson-Like scalings for turbulent velocity ($\sigma$--$L$) and gas density ($\rho$--$L$), respectively. These scalings rely on two key assumptions: a) molecular clouds maintain virial equilibrium across all scales, and b) there is an equivalence in the kinetic energy transfer rate between the virial velocity and turbulent velocity. Moreover, one can find that Eq.\,\ref{Larson Eq} serves as a specific solution the general scaling relations. However, the observed variations in the exponents of the $\sigma$--$L$ scaling  ($\sigma\propto L^{\beta}$; $\beta \sim 0.21-0.5$, e.g., \citealt{Lar81, Sol87, Cas95, Li23}) suggest the existence of possible multiple forms for Eq.\,(\ref{Larson Eq}). Therefore, these general scalings of both  $\sigma$--$L$ and $\rho$--$L$ need to be examined in realistic molecular clouds, especially in high-mass star formation clouds, which would help advance our understanding of the relative role of gravity, turbulence, and magnetic fields in high-mass star formation.

\section{Data set}\label{sec:datasets}
To explore scaling relations of both  $\sigma$--$L$ and $\rho$--$L$ in high-mass star formation regions, we compiled data from literature, including giant molecular clouds (GMCs), massive clumps, and massive cores. The dataset for different density structures within molecular clouds is summarized in Table\,\ref{tab: table1}. Note that our sample is not meant to be complete and unbiased (see more in Sect.\,4.4).  However, it encompasses a broad range of density structures spanning four orders of magnitude in spatial scale from 100\,pc to 0.01\,pc. This diversity allows us to conduct an overall analysis for the scalings of both  $\sigma$--$L$ and $\rho$--$L$.  We extracted essential parameters—such as mass, size, and velocity dispersion—from the references listed in the table. The determinations of these parameters were reexamined here prior to a systematic analysis in the following. 

\begin{center}
\begin{table}[]
    \centering
    \caption{A summary of the data and reference sources.}
    \begin{tabular}{cccccc}
    \hline
    \hline
         Density structure      & Mass tracer   & Kinematic tracer  & $n_{crit}$  & Data amount & Reference \\
         \hline
         Giant molecular cloud  & Molecule line & {\thco} (1-0)     & $\sim 10^3$ cm$^{-3}$    & 316     &\cite{Hey09} \\
          Massive clump          &  Continuum    & {\htcop} (1-0)    & $2\times10^5$ cm$^{-3}$  & 221     & \cite{Luo23} \\
          Massive clump and Core &  Continuum    & {\nth} (1-0)      & $3\times10^5$ cm$^{-3}$  & 44      & \cite{Oha16} \\
          Massive core           &  Continuum    & {\nth} (1-0)      & $3\times10^5$ cm$^{-3}$  & 11      & \cite{Per06} \\
          Massive core         & Continuum  & NH$_{3}$ (1-1)$/$(2-2) & $>10^4$ cm$^{-3}$         & 50      & \cite{Lu18} \\
          Massive core         &  Continuum  & N$_{2}$D$^{+}$ (1-0)  & $1.7\times10^6$ cm$^{-3}$ & 129     & \cite{Li23} \\
         \hline
    \end{tabular}
    \begin{flushleft}
{\bf Note:} $n_{crit}$ is the critical density for exciting a molecular line emission.
\end{flushleft}
    \label{tab: table1}
\end{table}
\end{center}

For the mass parameter ($M$), both molecular line and dust emission can be used for its estimate. On this regard, dust emission could be generally better than line emission since the latter could suffer from issues of unknown abundance and excitation conditions. In practice, for large-scale GMCs, where a reliable mass estimate from dust emission is not available, we relied on the mass results estimated by \citet{Hey09} using the {\thco} (1-0) molecular line emission. Note that due to the assumption of local thermodynamic equilibrium and variations in the abundance of {\thco} relative to molecular hydrogen, the mass of GMCs could be systematically uncertain by a factor of 2 to 3 \citep{Lis07, Hey09}. For the small-scale clumps and cores, we made use of the mass measurements from the dust continuum observations at far-IR to mm-wavelengths \citep{Per06, Oha16, Lu18, Li23, Luo23}. We rescaled the mass by using dust opacities following a common law, $\kappa_{\nu} = \kappa_{0}(\nu/\nu_0)^{\beta}$, where $\kappa_{0} = 0.1 \rm{cm^2 g^{-1}}$ at 1\,THz under a gas-to-dust mass ratio of 100 \citep{Bec90},  and  a fixed value of $\beta=1.5$ is adopted. Note that higher or lower dust opacities would result in a systematic uncertainty by a factor of up to two in dust mass measurements \egcite{Mot98, Kau13}.

For the radius parameter ($L$), various studies have employed different approaches. Some researchers define the source area within a contour of specific mass surface density or radiation intensity to determine the cloud radius \citep{Hey09, Luo23}. Alternatively, other studies utilize as the radius the Full Width at Half Maximum (FWHM) of the density distribution obtained from source identification algorithms  \citep{Per06, Oha16, Lu18, Li23}. In our analysis, we directly adopt the original measurements reported in the literature. Specifically, the median sizes for Giant Molecular Clouds (GMCs), clumps, and cores are 18.7 pc, 0.82 pc, and 0.04 pc, respectively. These values align with typical source sizes corresponding to various cloud density structures \citep{Bal20}. It is worth noting that the $L$-related parameters investigated here include mass ($M$), velocity dispersion ($\sigma$), and the virial parameter ($\alpha$). The third parameter is derived from the first two measured over the size $L$. Thus, analyzing the relationship between these parameters and size, rather than their absolute values, could not be affected by different approaches to measuring the size of density structures. In this sense, adopting the original measurements of the radius $L$ from the literature is sufficient for our following analysis.

Regarding the velocity dispersion parameter ($\sigma$), different molecular tracers were utilized for estimating velocity dispersion in the literature considered here. For large-scale, relatively low-density GMCs, where observations of high critical density tracers (e.g., {\nth} (1-0)) are unavailable primarily due to their difficulty of being excited in low density environments, the low critical density tracer {\thco} (1-0) line could be the best probe for kinematics of large-scale diffuse clouds, and thus was used to calculate the $\sigma$ parameter. For small-scale and relatively dense clumps and cores, high critical density line tracers including \htcop~(1-0), {\nth} (1-0), NH$_3$ (1-1) and (2-2), and N$_{2}$D$^{+}$ (1-0) were adopted, as summarized in Table\,\ref{tab: table1}. Most of these species are insensitive to the depletion effect due to freeze-out of molecules onto dust grains in dense environments \egcite{Cas99, Taf04, Ber07}, and are thus suitable for tracing the kinematics of dense clumps and cores.

With the related parameters determined (see above), we then calculated the mass density by $\rho = 3M/4\pi L^{3}$ and the virial parameter by {\avir} $= 5\sigma^{2}L/GM$ \citep{Ber92,Kau13}.  This definition of $\alpha_{\text{vir}}$ can be related to $a\frac{E_{\rm kin}}{|W_{\rm pot}|}$ \citep{Ber92}, where $E_{\rm kin}$ and $W_{\rm pot}$ represent the kinetic and gravitational potential energy, respectively. The coefficient $a$ accounts for non-homogeneous and non-spherical density distributions \egcite{Ber92, Mck99}. Recent observations have revealed that density structures can be inhomogeneous, as seen for example in high-mass star formation clumps with density gradients \egcite{Lin22, Gie23}. In a virial equilibrium condition ($2E_{\rm kin}\sim|W_{\rm pot}|$), the density structures have a critical virial parameter $\alpha_{\text{cr}}$, which corresponds to the $a$ coefficient as demonstrated by \citet{Kau13}, namely, $\alpha_{\text{cr}}=a$. Additionally, \citet{Kau13} evaluated the coefficient $a$ across a wide range of cloud shapes and density gradients (see their Appendix A), obtaining an approximately constant value of $2\pm1$.

In addition, the turbulent ram pressure and gravity pressure are determined by $P_{\mathrm{turbulent}} = \rho \sigma^{2}/3$ and $P_{\mathrm{gravity}} = G \rho^{2} {L}^{2}/\pi$ \citep{Li17}. The kinetic energy transfer rate per unit volume was derived via $\epsilon_{k} = \rho \sigma^{3}/L$ \citep{Kri07}.

\begin{figure}
\centering
\includegraphics[width=4.0 in]{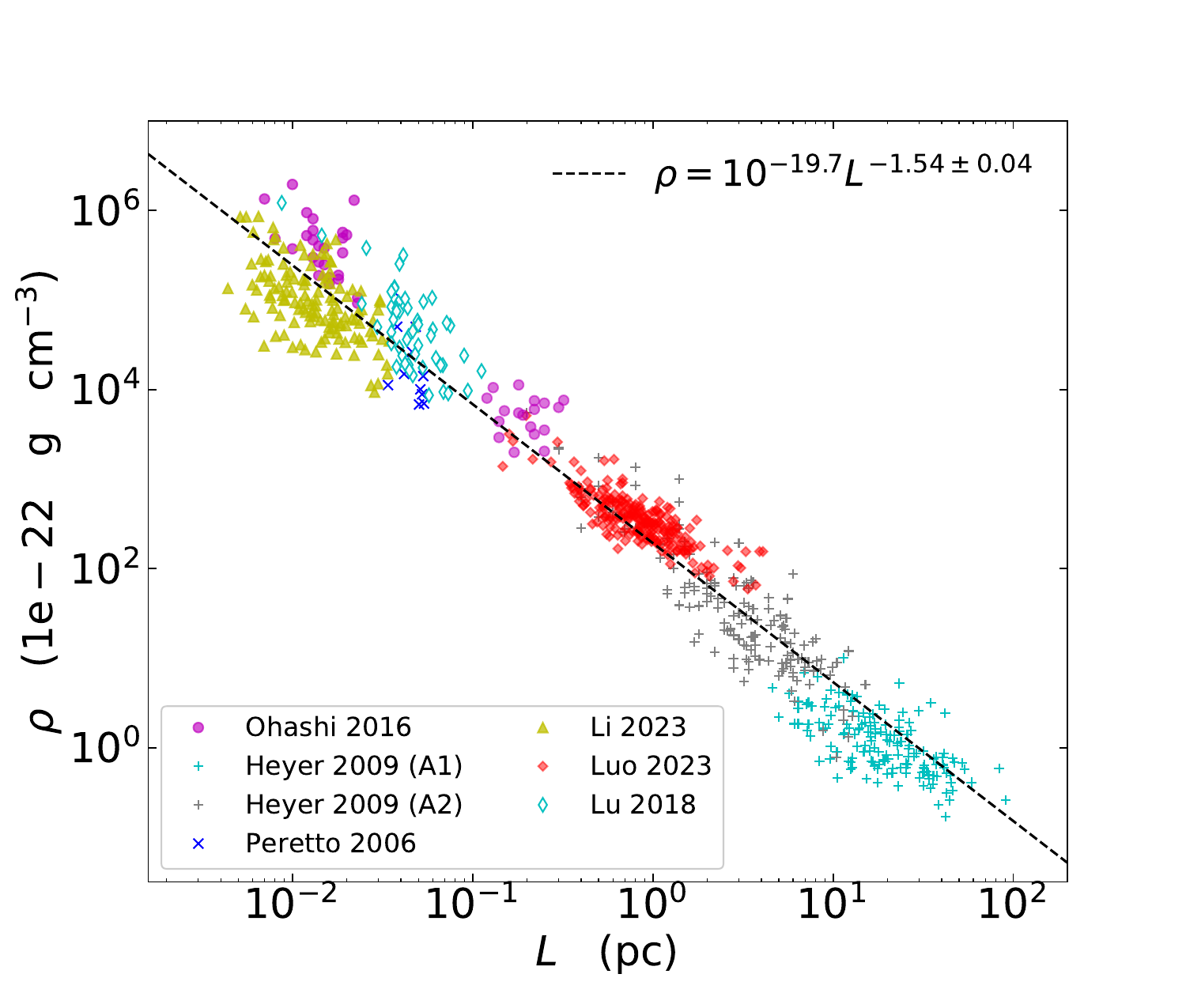} 
\caption{Density--size scaling across multiscales. Various symbols represent the data from different literature (see Table\,\ref{tab: table1}). The dashed line represents a linear fit in log-log space using the robust least squares regression (Huber regression) fitting approach. 
} 
\label{fig:rho-r fig}
\end{figure}

\section{scaling relations} \label{sec: virialized analysis}
\subsection{Scalings of both $\sigma$--$L$ and $\rho$--$L$}\label{subsec: subsec1}
Figure\,\ref{fig:rho-r fig} shows the density--size scaling (i.e., density profile) for our sample across multiscales. The density of hierarchical structures of MCs follows the relation $\rho = \rho_0 (L/\mathrm{pc})^{-p}$ with $\rho_0 = 10^{-19.7}\,\mathrm{g\, cm^{-3}}$, and $p = 1.54 \pm 0.04$. Plugging  this $\rho$--$L$ relation (i.e., $\rho \propto L^{-1.54}$) into Eq.\,\ref{virial v equation 0}, we can derive the scaling of velocity dispersion--size ($\sigma$--$L$)
\begin{align}
    \sigma \propto L^{0.23}.\label{virial v equation}
\end{align}

\begin{figure}
\centering
\includegraphics[width=4.0 in]{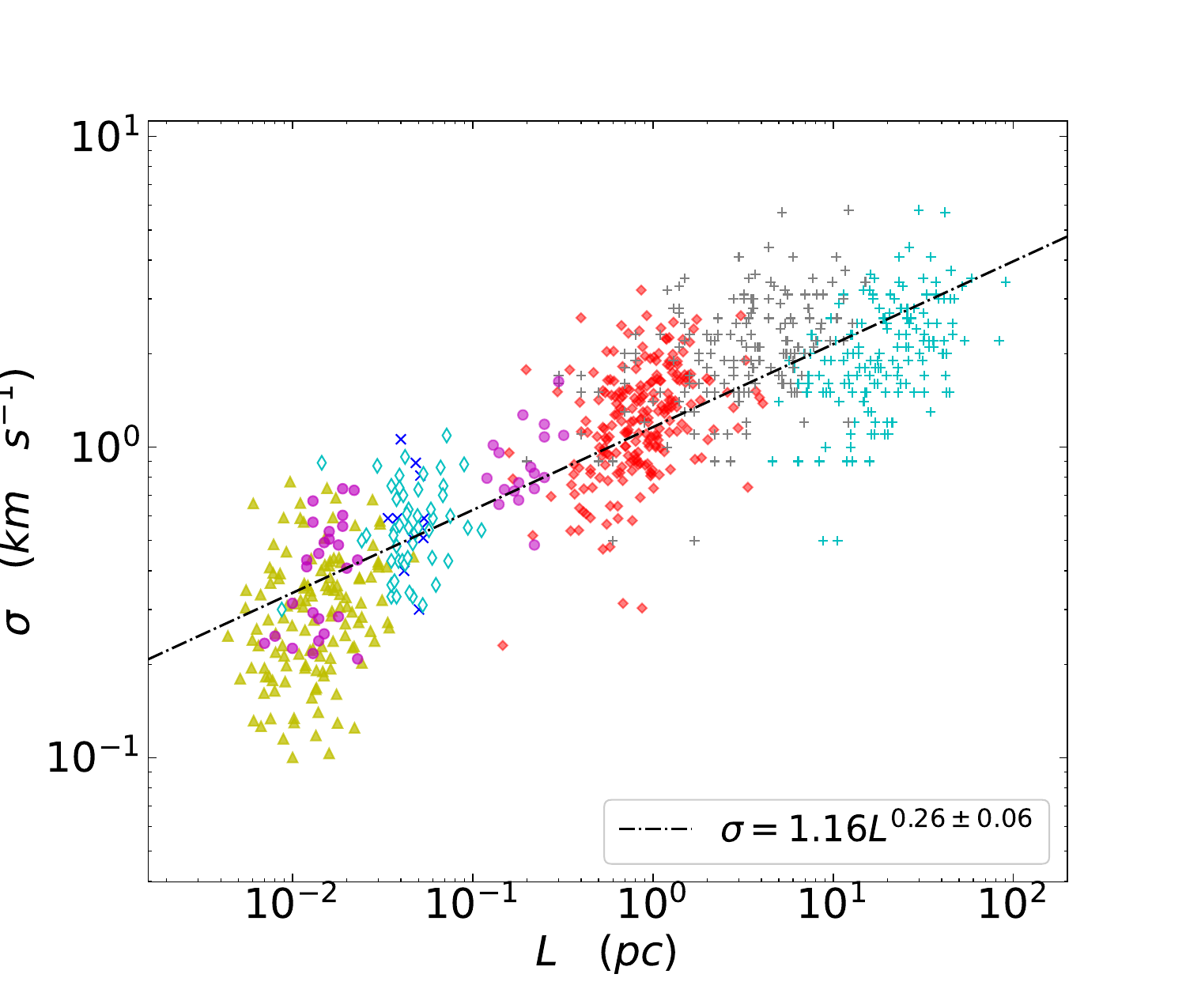} 
\caption{Same as Fig.\,\ref{fig:rho-r fig} but for the velocity dispersion--size scaling.}
\label{fig:l-r fig}
\end{figure}

Likewise, Fig.\,\ref{fig:l-r fig} displays the $\sigma$--$L$ scaling of our sample across multiscales. It shows that the observed scaling can be characterized as $\sigma = \sigma_0 (L/\mathrm{pc})^{\beta}$, where $\sigma_0 = 1.16$\,{\vel}, and $\beta = 0.26 \pm 0.04$. This scaling exponent ($\beta=0.26$) is much shallower compared to those previously reported (e.g., $\beta=0.38$ \citealt{Lar81} and $\beta=0.5$ \citealt{Sol87,Hey04}). However, the $\beta = 0.26$ observed here aligns with that in Eq. \ref{virial v equation} which relies on the observed $\rho$--$L$ scaling along with a virial equilibrium assumption.  This alignment indicates that from a global view high-mass star formation regions investigated here could maintain virial equilibrium by the competition between gravity and turbulence over multiscales from clouds down to cores.

\subsection{Virial parameter distribution}\label{subsec: subsec2}

\begin{figure}
\centering
\includegraphics[width=4.0 in]{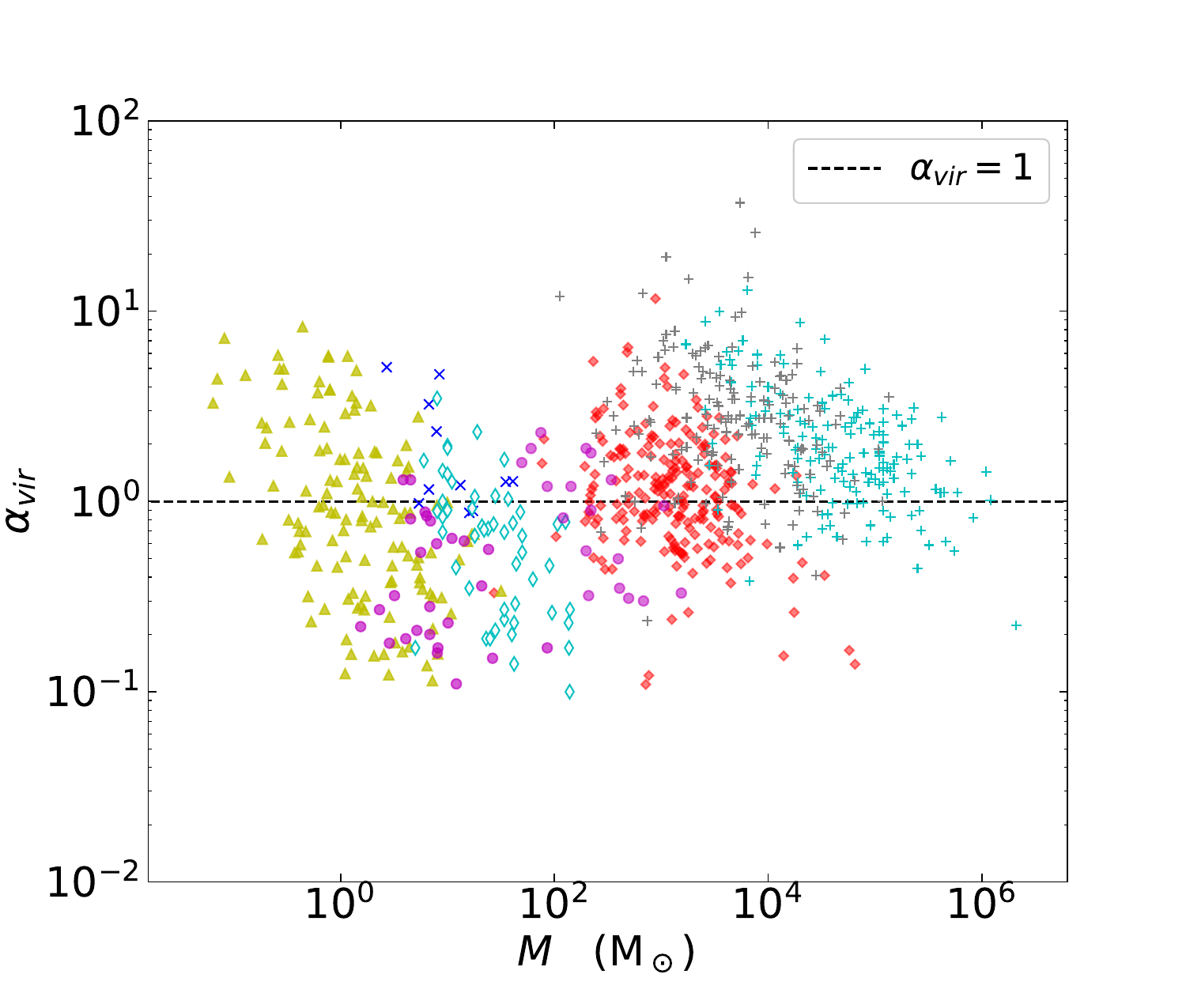} 
\caption{Virial parameter as the function of mass. Various symbols are the same as those in Figure\,\ref{fig:rho-r fig}. The dashed line indicates the unity.}
\label{fig:virial-m fig}
\end{figure}

The virial parameter can be used to gauge the virial state of molecular clouds. A molecular cloud keeps in virial equilibrium with the virial parameter {\avir} = $\alpha_{\rm cr}$. If we plug the observed scalings of both $\sigma$--$L$ and $\rho$--$L$  (i.e.,  $\rho \propto L^{-1.54}$, $\sigma \propto L^{0.26}$, respectively) into $\alpha_{\rm vir}=5\sigma^2L/GM$ with $M\propto\rho L^3$, we can derive the relation between {\avir} and mass as
\begin{align}
\alpha_{\mathrm{vir}} \propto M^{0.04}. 
\end{align}
 That is, the virial parameter remains approximately constant, irrespective of the mass of the density structures. This result can also be seen in Fig.\,\ref{fig:virial-m fig}, where we present {\avir} as a function of mass. Across all investigated density scales, the overall {\avir}$-M$ distribution is nearly flat, with {\avir} values centered around unity, close to $\alpha_{\rm cr}$, albeit with significant scatter on each individual scale (e.g., cores). The same result has been reported by \citet{Kau13}, who carefully calculated the parameters of both the virial and the mass using a common standardized method to alleviate their calculation uncertainties as much as possible. 
This result confirms the validity of the assumption of global virial equilibrium we made before, regulated by the interplay between gravity and turbulence, across multiple scales, from large-scale molecular clouds to small-scale cores.

\begin{figure}
\centering
\includegraphics[width=4.0 in]{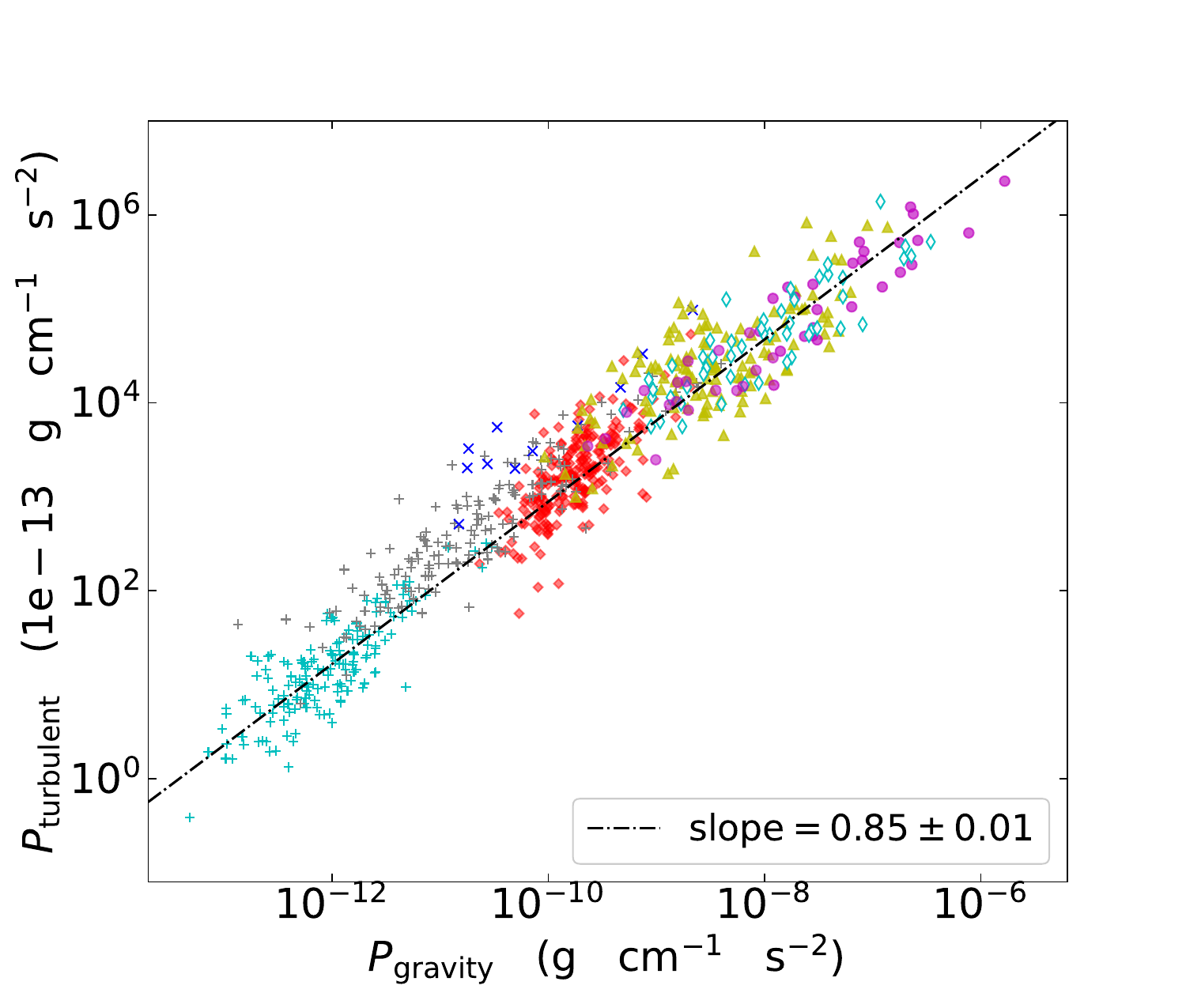} 
\caption{Comparison between turbulent and self-gravity pressures across multiscales. The dash dot line stands for a robust linear regression fitting in log-log space.  Other symbols are the same as those in Fig.\,\ref{fig:rho-r fig}.}
\label{fig:p fig}
\end{figure}

Moreover, in systems with supersonic turbulence maintaining virial equilibrium, the ram pressure from internal turbulence is thought to be comparable to self-gravity pressure \citep{Li17}. Figure\,\ref{fig:p fig} compares these pressures, showing a strong resemblance (Pearson's coefficient of 0.97) across all hierarchical density structures. This suggests again that high-mass star formation regions investigated here could maintain virial equilibrium across scales, from molecular clouds to cores. Note that the turbulent ram pressure is slightly lower than the self-gravity pressure (power law exponent of 0.85). This implies a slow gravitational contraction being at work across multiscales from clouds to cores,  akin to a quasi-static process.

For the large scatter (up to a factor of 10) around unity in the virial distribution on each individual density scale, it can correspond to a decreasing trend on each scale (see Fig.\,\ref{fig:virial-m fig}). We assume this decreasing trend, also observed in other studies \citep{Zha16,Tra18,Li20}, is likely due to the dynamical evolution of individual density scales, which oscillate between non-virial and virial equilibrium states. During this evolution, local dynamical states may deviate from the global virial equilibrium due to local fluctuations, such as stellar feedback like winds and outflows \citep{Mat02, Kle10, Gol11}. This hypothesis warrants further investigation through both theoretical and observational studies.

\subsection{Turbulent energy spectrum}\label{subsec3}
Turbulence in the interstellar medium (ISM) is predicted in theory to exhibit two modes. That is, the Kolmogorov-like incompressible mode \citep{Lar81} is characterized by an energy spectrum $E(k) \propto k^{-5/3}$ and a velocity scaling $\sigma \propto L^{1/3}$, while the Burgers-like compressible mode \citep{Li16, Li17} is described by an energy spectrum $E(k) \propto k^{-2}$ and a velocity scaling $\sigma \propto L^{1/2}$. Here, $k$ is the wavenumber, inversely proportional to spatial scale ($k \propto L^{-1}$). Our observed velocity scaling $\sigma \propto L^{0.26} \simeq L^{1/4}$ (see Fig.\,2) does not align with either mode, which can also be reflected from the turbulent energy spectrum as discussed below. 

Turbulence involves the transfer and dissipation of kinetic energy. For the turbulent interstellar medium, \cite{Fle96} proposed that the turbulent kinetic energy transfer rate per unit volume, $\epsilon_{k} = \rho \sigma^3 /L$, is invariant. This has been reproduced in simulations of large-scale three-dimensional isothermal supersonic Euler turbulent fluids \citep{Kri07,Hen12}.

\begin{figure}
\centering
\includegraphics[width=4.0 in]{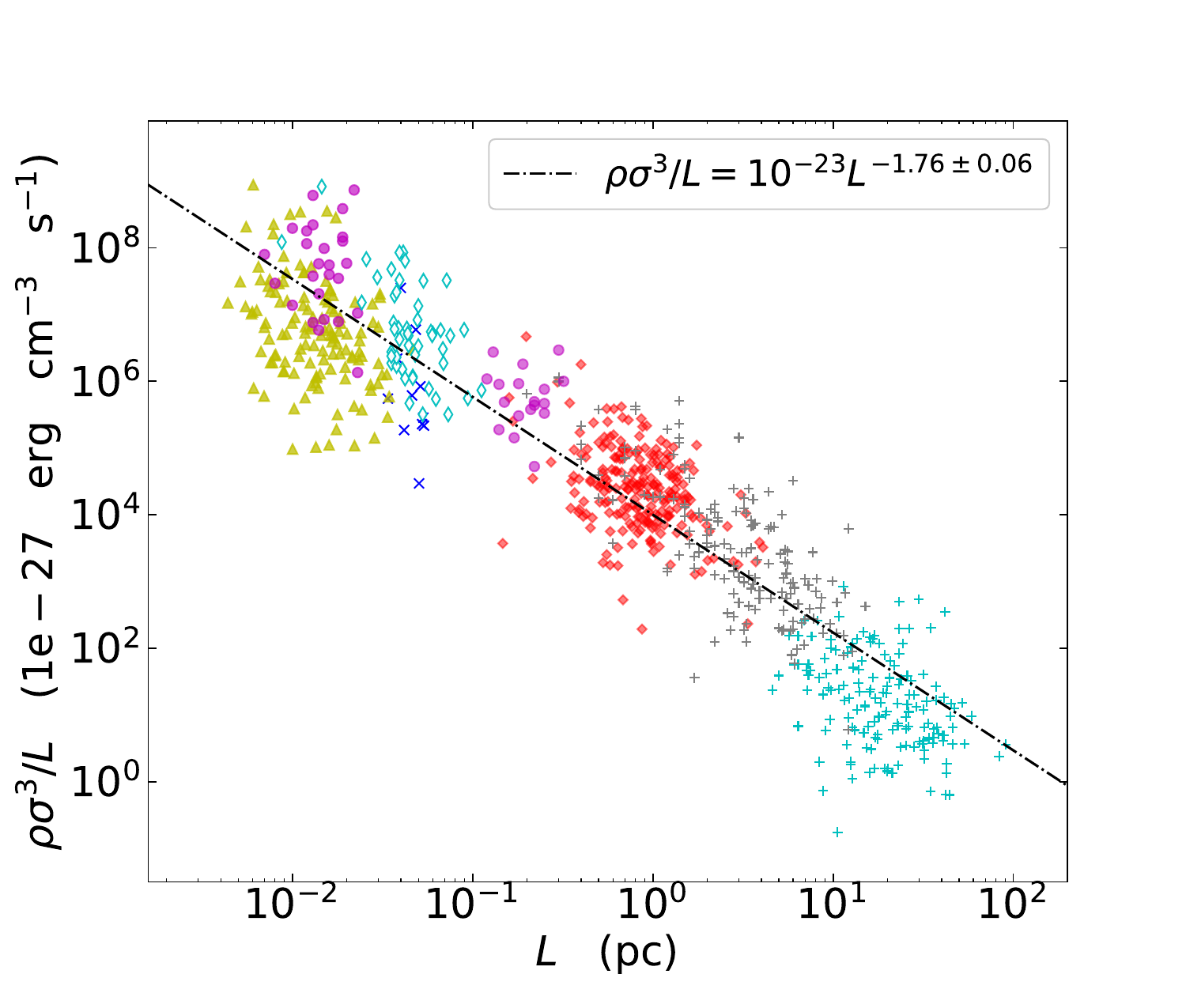} 
\caption{Kinetic transfer rate per unit volume as the function of scales. Other symbols are the same as those in Fig.\,\ref{fig:rho-r fig}.}
\label{fig:kt-r fig}
\end{figure}

However, we present in Fig.\,\ref{fig:kt-r fig} an observed relationship of $\epsilon_{k}$ as a function of scale, different from the predicted constant value.
The $\epsilon_{k}$ parameter is characterized as 
\begin{align}
    \epsilon_{k} = \rho \sigma^3 /L = \epsilon_0 (L/\mathrm{pc})^{-\gamma},
\end{align}
where $\epsilon_0 = 10^{-23} \,\mathrm{erg\,cm^{-3}\,s^{-1}}$, and $\gamma = 1.76 \pm 0.06$, both derived from Fig.\,\ref{fig:kt-r fig}. 
Combining the density profile $\rho \propto L^{-p}$ and the $\epsilon_{k} \propto L^{-\gamma}$, we can re-express the velocity dispersion--size scaling as
\begin{align}
    \sigma \propto L^{(1+p-\gamma)/3},
\end{align}
and accordingly 
\begin{align}
    \sigma \propto k^{-(1+p-\gamma)/3}.\label{v equation}
\end{align}

According to the turbulent energy spectrum function \citep{Fle96} 

\begin{align}
    E(k) = \frac{1}{2}\frac{d \sigma^{2}}{dk},\label{E equation}
\end{align}
substituting Eq.\,\ref{v equation} into Eq.\,\ref{E equation}, we reach to
\begin{align}
    E(k) \propto k^{-(5+2p-2\gamma)/3}.
\end{align}
Recalling the observed $p = 1.54$ (see Fig.\,1) and $\gamma = 1.76$ (see Fig.\,5), we find
\begin{align}
    E(k) \propto k^{-1.52}.
\end{align}
The energy spectrum index of -1.52 reveals a shallower energy spectrum compared to the energy spectrum of both compressible and incompressible turbulence modes. Nevertheless, we find the index of -1.52 consistent with that of energy spectrum for hydromagnetic turbulence \egcite{Iro64, Kra65}, which is characterized as $E(k) \propto k^{-3/2}$.
This consistency suggests that turbulence and magnetic fields may collaboratively counteract gravity, thereby regulating the dynamics of MCs and maintaining their global dynamical equilibrium. This may correspond to a widespread presence of turbulence and magnetic fields in the interstellar medium \citep{Fri95, Elm04, Hen12, Fed13,Li13, Pla16}.

In addition, by definition, $\alpha<\alpha_{\rm cr}$ implies that a cloud structure or fragment is susceptible to collapse. However, \cite{Kau13} argued that fragments with $\alpha<\alpha_{\rm cr}$ are unlikely to be in a state of collapse, a notion previously realized by \cite{Lar81} and \cite{Bal06}. If true, the most straightforward explanation for such fragments would be that they are supported against collapse by significant magnetic fields. This accordingly suggests that magnetic fields, in addition to turbulence, may play a role in regulating cloud dynamics, as revealed by the observed turbulent energy spectrum (see Fig.\,\ref{fig:kt-r fig}) likely responsible for magneto-hydrodynamical turbulence.

It is worth noting that the discussions above require further confirmation through direct measurements of magnetic fields. This is because the energy spectrum index of -1.52 observed here could also arise from other factors, such as large-scale forcing or anisotropic turbulent motions \egcite{Val18}.

\subsection{Caveats}
Our discussions on the $\sigma$--$L$ and $\rho$--$L$ scalings as well as the associated turbulent energy spectrum may be influenced by observational biases, as our data integrate different literature involving various telescopes and molecular tracers. This influence may be mitigated somehow by re-examining some parameters (e.g., mass, size, and velocity dispersion) using a consistent approach as described in Sect.\,3 to re-examine the determinations of the related parameters (e.g., mass, velocity dispersion). Given these potential biases, we shift our focus on global scalings across multiscales only, rather than local counterparts on each individual scale (e.g., cores or clumps). In addition to greater impact of the observational biases on local scaling analysis, and the limited dynamical range on each individual scale does not allow the same analysis as robustly as possible. Overall, for more robust discussions on the $\sigma$--$L$ and $\rho$--$L$ scalings as well as the associated turbulent energy spectrum, future work should aim for consistent, multiscale kinematic observations of molecular clouds using the same telescope and molecular tracer.

\subsection{Implication on high-mass star formation}
As discussed earlier, gravity, turbulence, and even possibly magnetic fields collectively regulate the dynamics of molecular clouds. This could provide a global implication on the latest theoretical models of high-mass star formation, such as GHC and I2. Both models posit that high-mass star formation in molecular clouds is a multiscale process involving fragmentation and mass accretion at various density scales. 

While both GHC and I2 models agree that gravity drives mass accretion on smaller scales, such as cores/clumps and filaments, they diverge on larger scales. The GHC model suggests that gravity primarily drives hierarchical mass accretion across all cloud density scales \citep{Vaz19}, while the I2 model proposes that turbulence regulates mass inflow and accretion on large-scale clouds, with self-gravity control assumed at smaller scales \citep{Pad20}. 

Our observations, including the $\sigma$--$L$ and $\rho$--$L$ scalings, the {\avir} distribution centered around unity regardless of density scales, the balance between turbulent pressure and self-gravity pressure, and the turbulent energy spectrum, all suggest that the multiscale process of high-mass star formation could be regulated by gravity, turbulence, and possibly magnetic fields collectively. Therefore, current simulations of high-mass star formation would need to incorporate these major factors for a comprehensive understanding of the multiscale scenario of high-mass star formation.

\section{Summary and Conclusions}\label{sec:conclusions}
We have explored the $\sigma$--$L$ and $\rho$--$L$ scalings and the associated turbulent energy spectrum using a large data sample from various sources over multiscales from 0.01 pc to 100 pc. This sample spans different hierarchical density structures in high-mass star formation clouds, from giant molecular clouds to clumps and dense cores. Our findings suggest that gravity, turbulence, and  possibly magnetic fields all together could play key roles in regulating the dynamics of molecular clouds and high-mass star formation therein.

Our major findings include the $\sigma$--$L$ and $\rho$--$L$ scalings, $\sigma \propto L^{0.26}$ and $\rho \propto L^{-1.54}$, which can lead to a state of virial equilibrium. We also observe a nearly flat {\avir}$-M$ distribution across all density scales, with {\avir} values centered around unity, despite significant scatter on each scale. This supports the idea of a global equilibrium across multiple scales, maintained by the balance between gravity and turbulence. This is further reinforced by the observed balance between turbulent pressure and self-gravity pressure. Our analysis of the turbulent energy spectrum, based on the $\sigma$--$L$ and $\rho$--$L$ scalings, reveals a characteristic $E(k) \propto k^{-1.52}$, possibly consistent with the prediction of magneto-hydrodynamical turbulence. This consistency suggests that magnetic fields, alongside turbulence, may contribute to the regulation of cloud dynamics. All of these findings warrants further confirmation through future multiscale kinematic observations of molecular clouds with uniform observing settings, for example using the same telescope in the same kinematic tracer. 

\begin{acknowledgements}
We thank the anonymous referee for comments and suggestions that greatly improved the quality of this paper. This work has been supported by the National Key R\&D Program of China (No.\,2022YFA1603101). H.-L. Liu is supported by National Natural Science Foundation of China (NSFC) through the grant No.\,12103045, by Yunnan Fundamental Research Project (grant No.\,202301AT070118, 202401AS070121), and by Xingdian Talent Support Plan -- Youth Project. G.-X. Li is supported by NSFC under No.\,12033005.
\end{acknowledgements}

\bibliographystyle{raa}
\bibliography{RAA}

\label{lastpage}

\end{document}